\documentclass[10pt,letterpaper,onecolumn]{article}

\usepackage[letterpaper,top=1in,bottom=1in,left=0.75in,right=0.75in]{geometry}

\usepackage{times}

\usepackage{amsmath}
\usepackage{graphicx}
\usepackage{url}

\title{Scoring Azure permissions with metric spaces}
\author{
        Christophe Parisel\thanks{Email: ch.parisel@gmail.com}
}

\begin{document}

\maketitle

\begin{abstract}
In this work, we introduce two complementary metrics for quantifying and scoring privilege risk in Microsoft Azure. In the Control Plane, we define the \emph{WAR distance}, a superincreasing distance over Write, Action, and Read control permissions, which yields a total ordering of principals by their configuration power. 

In the Data Plane, we present a \emph{blast radius distance} for measuring the maximum breadth of data exfiltration and forgery, leveraging the natural ultrametric induced by Azure Tenants' clustering hierarchy.

Collectively, these distances offer a unified framework for proactive IAM analysis, ranking, lifecycle monitoring, and least privilege enforcement.
\end{abstract}

\section{Introduction}

In modern cloud environments like Microsoft Azure, overseeing the security of privileged access is a complex and critical challenge, 
especially for \emph{Non Human Identities} (NHIs), such as Application SPNs and Managed Identities. As organizations move to the cloud, the need for robust IAM risk-measurement systems has become more pressing. One of the most significant risks is the potential for excessive or misconfigured privileges granted to NHIs aside of 
the standard groups membership process that human principals typically follow, which can lead to unauthorized data exfiltration, integrity breaches, or unintended privilege escalation.

Azure provides fine-grained access control through role-based access control (RBAC) \cite{msft_rbac,fer}, allowing principals (users, service principals, or managed identities) to be assigned permissions on resources across various scopes such as tenants, management groups (MG), subscriptions, resource groups (RG), and individual resources. However, understanding the exact risk posed by these permissions, particularly when they are distributed across different scopes and types, is a non-trivial problem.

This paper attempts to solve that problem leveraging natural metric spaces found in Azure. We begin with an overview of the WAR distance metric, explaining how it combines scope and permission type to measure control-plane privilege. We then present the data-plane ultrametric for assessing risks related to data exfiltration and forgery.

\section{Measuring Control Plane Privilege: The WAR Distance}

\subsection{Motivation and Problem Statement}

In Microsoft Azure's Role-Based Access Control (RBAC) system, principals (users, service principals, or managed identities) are granted permissions through role assignments that specify both \emph{what} operations can be performed and \emph{where} they can be performed. This creates a complex privilege landscape where understanding the relative risk of different permission combinations is non-trivial.

Consider two principals: one with \texttt{Microsoft.Compute/virtualMachines/read} at the tenant level, and another with \texttt{Microsoft.*/write} at a single resource group. Which poses greater risk? Traditional approaches struggle with such comparisons because they lack a principled method for combining scope breadth with operation criticality.

We introduce the WAR (Write/Action/Read) distance metric to address this challenge. Our approach leverages the mathematical properties of super-increasing sequences to create a unified scoring system that enables total ordering of principals by their control plane privilege strength.

\subsection{Mathematical Foundation: Super-increasing Sequences}

\subsubsection{Super-increasing Sequence}
\textbf{Definition:} A sequence $s_1, s_2, \ldots, s_n$ is super-increasing if each element is greater than the sum of all preceding elements:
$$s_i > \sum_{j=1}^{i-1} s_j \quad \text{for all } i > 1$$

Super-increasing sequences possess a crucial property: they enable unique representation of any linear combination with binary coefficients.

\medskip

\textbf{Lemma (Unique Decomposition Property):}
Let $S = \{s_1, s_2, \ldots, s_n\}$ be a super-increasing sequence. For any subset $T \subseteq S$, the sum $\sum_{s \in T} s$ uniquely determines $T$.

\medskip

\textbf{Proof}
Suppose two distinct subsets $T_1, T_2 \subseteq S$ yield the same sum. Let $s_k$ be the largest element in the symmetric difference $T_1 \triangle T_2$. Without loss of generality, assume $s_k \in T_1 \setminus T_2$. Then:
$$\sum_{s \in T_1} s \geq s_k > \sum_{j=1}^{k-1} s_j$$

$T_2$ can at most include all elements smaller than $s_k$, hence:

$$ \sum_{j=1}^{k-1} s_j \geq \sum_{s \in T_2} s$$

Finally,

$$\sum_{s \in T_1} s > \sum_{s \in T_2}$$

This contradicts the assumption that the sums are equal, proving uniqueness.

\medskip

This property is fundamental to our metric construction, as it ensures that privilege scores can be unambiguously decomposed into their constituent components.

\subsection{The WAR Distance Model}

\subsubsection{Scope and Permission Type Hierarchies}

We model Azure's control plane permissions as elements of the Cartesian product of two ordered spaces:

\textbf{Scope Space} (by increasing specificity):
$$\text{Tenant} \prec \text{Management Group} \prec \text{Subscription} \prec \text{Resource Group} \prec \text{Resource} \prec \text{Sub-resource}$$

\textbf{Permission Type Space} (by increasing risk):
$$\text{Read} \prec \text{Action} \prec \text{Write} \prec \text{Wildcard}$$

\subsubsection{Weight Assignment via Super-increasing Sequences}

We construct three super-increasing sequences corresponding to our permission types:

\begin{align}
W &= \{900, 800, 700, 600, 500, 400, 300, 200, 100\} \quad \text{(Write weights)} \\
A &= \{90, 80, 70, 60, 40, 20\} \quad \text{(Action weights)} \\
R &= \{9, 8, 7, 6, 4, 2\} \quad \text{(Read weights)}
\end{align}

These sequences satisfy the super-increasing property both within and across permission types:
\begin{align}
\min(W) &= 100 > 90 + 9 = \max(A) + \max(R) \\
\min(A) &= 20 > 9 = \max(R)
\end{align}

The extended Write sequence includes superadmin weights (900, 800, 700) for wildcard permissions at tenant, management group, and subscription levels respectively, reflecting their elevated privilege.

\subsubsection{Principal Representation and Distance Function}

\textbf{Definition (Principal Tuple):}
A principal $P$ with control plane permissions is represented as a tuple:
$$P = (w, a, r)$$
where $w \in W$, $a \in A$, and $r \in R$ represent the highest-weighted Write, Action, and Read permissions held by the principal across all scopes.

\medskip

\textbf{Definition (WAR Distance):}
For two principals $P_1 = (w_1, a_1, r_1)$ and $P_2 = (w_2, a_2, r_2)$, the WAR distance is:
$$d(P_1, P_2) = |w_1 + a_1 + r_1 - w_2 - a_2 - r_2|$$

\medskip

\textbf{Theorem:} The WAR Distance is a Metric \newline
The WAR distance function $d: P \times P \to R^+$ is a proper metric on the space of principals $\mathcal{P}$.

\medskip
\textbf{Proof}

We verify the four metric axioms:
\begin{enumerate}
\item \textbf{Non-negativity}: $d(P_1, P_2) = |w_1 + a_1 + r_1 - w_2 - a_2 - r_2| \geq 0$ by definition of absolute value.

\item \textbf{Symmetry}: $d(P_1, P_2) = |w_1 + a_1 + r_1 - w_2 - a_2 - r_2| = |w_2 + a_2 + r_2 - w_1 - a_1 - r_1| = d(P_2, P_1)$.

\item \textbf{Triangle inequality}: For any three principals $P_1, P_2, P_3$:
\begin{align}
d(P_1, P_3) &= |(w_1 + a_1 + r_1) - (w_3 + a_3 + r_3)| \\
&= |(w_1 + a_1 + r_1) - (w_2 + a_2 + r_2) + (w_2 + a_2 + r_2) - (w_3 + a_3 + r_3)| \\
&\leq |(w_1 + a_1 + r_1) - (w_2 + a_2 + r_2)| + |(w_2 + a_2 + r_2) - (w_3 + a_3 + r_3)| \\
&= d(P_1, P_2) + d(P_2, P_3)
\end{align}

\item \textbf{Identity of indiscernibles}: $d(P_1, P_2) = 0$ if and only if $w_1 + a_1 + r_1 = w_2 + a_2 + r_2$. By the unique decomposition lemma and the super-increasing property of our weight sequences, this occurs if and only if $(w_1, a_1, r_1) = (w_2, a_2, r_2)$, i.e., $P_1 = P_2$.
\end{enumerate}

\subsection{Privilege Strength and Principal Ranking}

\textbf{Definition (Privilege Norm):} \newline

For a principal $P = (w, a, r)$, we define its privilege strength as: $$\|P\| = w + a + r$$

\medskip

This norm, induced by the WAR distance from the null principal $O = (0, 0, 0)$, provides a scalar measure for ranking principals by their control plane privilege. The super-increasing property ensures that:

\begin{enumerate}
\item \textbf{Unambiguous ranking}: No two distinct privilege configurations yield the same norm.
\item \textbf{Hierarchical decomposition}: Given any norm value, we can uniquely determine the underlying permission structure.
\item \textbf{Additive interpretation}: Higher-scope permissions dominate lower-scope ones in the total score.
\end{enumerate}

\subsubsection{Norm Bounds and Interpretation}

The privilege norm ranges from 0 to 999:
\begin{itemize}
\item $\|P\|_{\min} = 0$: Null principal with no control plane permissions
\item $\|P\|_{\max} = 999 = 900 + 90 + 9$: Tenant-level superadmin with full wildcard, action, and read permissions
\end{itemize}

\medskip

Key reference points include:
\begin{itemize}
\item $\|P\| = 777$: Subscription-level superadmin
\item $\|P\| = 690$: Tenant-level read-only (900 for wildcard read at tenant level)
\item $\|P\| = 477$: Subscription-level full permissions without wildcard
\end{itemize}

\subsection{Handling Group Membership}

Real-world principals often inherit permissions through group membership. We extend our model to account for transitive privilege inheritance:

\medskip

\textbf{Definition (Effective Principal Tuple):} \newline
For a principal $P$ belonging to groups $G_1, G_2, \ldots, G_k$, the effective privilege tuple is:
$$P_{\text{eff}} = \left(\max_{i \in \{0, 1, \ldots, k\}} w_i, \max_{i \in \{0, 1, \ldots, k\}} a_i, \max_{i \in \{0, 1, \ldots, k\}} r_i\right)$$
where $(w_0, a_0, r_0)$ represents the principal's direct permissions and $(w_i, a_i, r_i)$ represents the permissions of group $G_i$.

\medskip

While this extension destroys the unique decomposition for individual privilege sources, max-based aggregation is a straightfoward way to reflect the reality that principals can exercise the highest privileges available through any of their group memberships.

\subsection{Operational Applications}

The WAR distance metric is fully implemented in silhouette\cite{silhouette}, an open-source tool. It enables several practical security applications:

\begin{description}
\item[Privilege Ranking] Sort principals by $\|P\|$ to identify the most privileged identities requiring attention.

\item[De-escalation Planning] For a principal with current norm $\|P_{\text{current}}\|$ and target norm $\|P_{\text{target}}\|$, the WAR distance $d(P_{\text{current}}, P_{\text{target}})$ quantifies the privilege reduction required.

\item[Anomaly Detection] Monitor $\|P(t)\|$ over time to detect unusual privilege escalations or configuration drift.

\item[Compliance Reporting] Use norm thresholds to enforce organizational privilege policies and generate compliance reports.
\end{description}

The mathematical rigor of the WAR distance, grounded in super-increasing sequences, ensures that these applications rest on a solid theoretical foundation while remaining computationally efficient and practically applicable for direct permission assignments. In the groups membership extension, the unique WAR decomposition is lost but the WAR norms remain comparable.

\newpage

\section{Measuring Data Plane Privilege: The Blast Radius Distance}

\subsection{Introduction}
In the context of data plane security, two fundamental risks dominate the threat landscape: \textbf{data exfiltration} and \textbf{data forgery}. These risks arise from overprivileged identities possessing \texttt{dataActions} on resources, often across organizational boundaries.
\paragraph{Data Exfiltration (Confidentiality).}
Read-level \texttt{dataActions} grant a principal the ability to access and extract sensitive information from Azure data resources. When such permissions are distributed across logically independent scopes—such as separate business units, regions, or environments, the risk of large-scale exfiltration escalates. In the event of a compromise, an attacker could harvest data across these boundaries, breaching confidentiality in a systemic fashion.
\paragraph{Data Forgery (Integrity).}
Write-level \texttt{dataActions} empower a principal to modify or overwrite existing data. This capability introduces a serious risk to data integrity: a compromised identity with write access could forge fraudulent data, inject malicious payloads, or corrupt operational information. The breadth of such permissions determines the potential impact of a forgery attack.

\subsection{Understanding Exfiltration And Forgery In The Azure Context}
Let's revisit permissions types and scope under the light of data plane security in Azure, and compare them with what we proposed in the control plane.

\subsubsection{Prioritization Of \texttt{dataActions} Permission Types}

Azure's data plane permissions comprise four primary types:

\begin{itemize}
    \item \textbf{Wildcard (\texttt{*})} permission grants unrestricted access to all data plane actions within the scope of the assigned role.
    \item \textbf{Write} permission enables creation, deletion, or modification of customer data.
    \item \textbf{Read} permission grants visibility into customer data.
    \item \textbf{Action} permission typically allows enumeration operations, such as listing blobs, keys, or containers.
\end{itemize}

Our model introduces a custom risk-driven prioritization, emphasizing the potential for data breach scenarios:

\begin{enumerate}
    \item \textbf{Read or Write} permissions are both high-risk, as they enable data exfiltration (confidentiality breach) and data forgery (integrity compromise), respectively.
    \item The combination of \textbf{Read and Write} permissions elevates the risk further, enabling simultaneous confidentiality and integrity attacks. This combination defines the highest data-centric privilege.
    \item \textbf{Action} permissions are comparatively low-risk in the data plane and are excluded from blast radius scoring due to their limited exploitability. They are still opening doors for reconnaissance, so they should not be ignored altogether. But as far as data security is concerned, they play a secondary role.
    \item \textbf{Wildcard} permissions are treated as functionally equivalent to the union of Read and Write, as the latter already imply maximal impact on both confidentiality and integrity.
\end{enumerate}

\subsubsection{Preservation Of Organizational Context}

To focus on business-impactful privilege boundaries, we intentionally collapse all scopes \emph{below} the subscription level. Whether a principal has access at the customer database, resource group, or entire subscription level, the risk to business-critical data remains fundamentally the same in the data plane.

In contrast, the hierarchy of Management Groups (MGs) is preserved \emph{as-is}, due to its organizational relevance. This structure encapsulates real-world data segmentation practices and is important in modeling blast radii:

\begin{itemize}
    \item \textbf{Production vs. Non-Production} – Customers frequently segment prod and non-prod environments via separate MGs. Cross-boundary access is a clear data risk and must be surfaced.
    \item \textbf{Geographical Boundaries} – Regional compliance requirements (e.g., GDPR, data residency laws) are often implemented using MG segmentation. Cross-region data access poses regulatory risks.
    \item \textbf{Business Units (BUs)} – Enterprises model internal departments or legal entities as nested MGs. In industries like finance, this ensures adherence to ``Chinese Wall'' policies. Access across BUs signifies a significant breach of data containment.
\end{itemize}

Preserving the MG hierarchy in our blast radius computation allows us to align privilege assessment with customer-specific risk profiles and real-world operational boundaries.

\subsubsection{Implementation}

The Blast radius is fully implemented in the silhouette\cite{silhouette} tool.

\subsubsection{Comparison With Control Plane Needs}

\paragraph{Control Plane Context Scoping}
In contrast to the data plane, control plane operations can have profound consequences even at fine-grained scopes such as individual resources or sub-resources. As such, the WAR metric captures permissions down to these granular levels. On the other hand, as control plane scopes widen, their marginal impact tends to plateau: once a principal can configure one Management Group (MG), they can likely affect others within the tenant in a similar fashion. Therefore, the WAR model simplifies the hierarchy by collapsing all MGs into a single abstract level, reflecting the observation that compromising one MG is operationally similar to compromising several within the same tenant.

\paragraph{Permission Types}
In the WAR model, permission types are ordered based on their potential to cause configuration damage or privilege escalation. The ordering is as follows:
\[
	\texttt{*} \geq \texttt{Write} > \texttt{Action} > \texttt{Read}
\]
This reflects the fact that \texttt{Write} permissions can be used to take over or reconfigure resources, while \texttt{Read} permissions alone pose relatively little configuration risk. 
Additionally, wildcard permissions (e.g., \texttt{*}) in the control plane are particularly hazardous (especially for scopes at or above the subscription level) because they enable arbitrary actions and may implicitly include future actions not currently defined and potentially catastrophic breaking changes. 

\paragraph{Threat focus}
Data protection focuses on direct threats to data confidentiality (exfiltration) and integrity (forgery). In contrast, control plane risks often center on misconfigurations, such as disabling audit logs or weakening encryption policies.

\paragraph{Scope Modeling} 
In the data plane, we abstract away fine-grained resource details and evaluate risk at the subscription level. This choice reflects operational realities: access to a resource or its containing subscription typically results in equivalent data exposure. However, we retain the hierarchical structure of Management Groups to capture organizational segmentation (e.g., production vs. non-production, business units, or geographic regions).

\paragraph{Permission Prioritization} 
Compare permissions prioritization in the control plane WAR model:
\[
\texttt{*} > \texttt{Write} > \texttt{Action} > \texttt{Read}
\]
with our data plane model, where confidentiality and integrity are equally critical:
\[
\texttt{*} = (\texttt{Write} \text{ and } \texttt{Read}) > (\texttt{Write} \text{ or } \texttt{Read}) > \texttt{Action}
\]

\paragraph{Risk Implications} 
A high data plane blast radius signals that a principal holds read or write access across multiple independent organizational zones. This fragmentation amplifies the risk of data exfiltration or tampering, distinguishing it from the predominantly configuration-centric risks of the control plane.

\subsection{The Blast Radius Distance}
To capture and quantify data exfiltration and data forgery, we introduce a new purpose-built metric in Azure: the \emph{Blast Radius Distance}. 
This distance measures how far a principal’s data privileges extend across the Azure resource hierarchy. It is designed to reflect both the depth and dispersion of sensitive \texttt{read} and \texttt{write} permissions, enabling a structured assessment of data risk exposure for any given identity.

\subsubsection{The Distance Model}
We leverage the natural ultrametric distance \cite{ultrametric,intro} between two dataActions permissions $p_1$ and $p_2$ of the principal in the Azure Tenant hierarchy: 
\begin{itemize}
\item Depth 0: Tenant (Root) 
\item Depth 1+: Management Groups (MGs), up to 6 officially 
\item Final depth ``d'': Subscription (treated as the leaf, with all sub-resources collapsed upward)
\end{itemize}
So depth ranges from 0 to a maximum of 7. 

\medskip
\noindent

To account for groups membership, we must consider all pairs of dataActions permissions inherited from the groups the 
principal belongs to.

\subsubsection{The Distance Function}
For any two scopes $s_1$ and $s_2$ of permissions $p_1$ and $p_2$, let d be the depth of their Lowest Common Ancestor (LCA). We define the base ultrametric distance as \\

$\mathrm{D}(s_1, s_2) = \frac{1}{2^{2d+1}}$

\medskip
\noindent

\par
The base distance ranges from 0.0 (no data plane rights or infinitely small blast radius) to 1.0 (Tenant wide radius) 
It defines measurements and holes in the $\frac{1}{2^{n}}$ series: measurements appear at odd powers in the series, and holes at even ones.

\medskip
\noindent

To reflect the permissions ordering, we ignore all permission pairs containing an `Action'. 
For the remaining pairs, we consider the shallowest scope depth of a pair of permissions as $min(s_1,s_2)$ and define an `impact' coefficient:
\begin{itemize}
\item Impact = 2 if the pair contains a wildcard at shallowest scope depth, or if at least  
`Write' and `Read' atomic permissions are assigned to the principal at shallowest scope depth. 
\item Impact = 1 if the `Read' or `Write' permission is assigned to the principal at shallowest scope depth, 
but not both simultaneously at this depth.
\end{itemize}
The final distance is: 
$\mathrm{\delta}(s_1, s_2) = impact(p_1,p_2) \cdot D(s_1,s_2)$
\par

\medskip
\noindent

Concretely, for high-risk permissions, since impact=2 the final distance is 
$\frac{1}{2^{2d}}$ whereas for lower-risk permissions since impact=1 the final distance is $\frac{1}{2^{2d+1}}$ 
\par

This nuance defines a hierarchy within the hierarchy: it fills in holes in the series in an orderly fashion: for identical scopes, high risk
permissions at even locations have a higher measurement than lower risk ones at odd locations.

\subsubsection{Formalization Of The Blast Radius}

We define the data plane blast radius by leveraging a fundamental property of ultrametric distances: the strong triangle inequality, given as:
\[
\text{distance}(a, c) \leq \max\big(\text{distance}(a, b), \text{distance}(b, c)\big)
\]

\medskip
\noindent
\textbf{Definition:} For a principal $P$ with a set of data plane permissions $\mathcal{P} = \{p_0, \ldots, p_n\}$ with scopes $\mathcal{S} = \{s_0, \ldots, s_n\}$, we define the blast radius as:
\[
	\max_{i \neq j \in \mathcal{S}} \delta(s_i, s_j) \text{ if } |\mathcal{P}| > 1,
\]
\[
	\frac{impact(p_0,p_0)}{2^{2d+1}} \text{ if } |\mathcal{P}| = 1 \text{ (singleton case) }
\] 

\medskip

\noindent
When $|\mathcal{P}| > 1$, this value represents the \textbf{diameter} of $\mathcal{P}$: how far apart the most distant permissions are, organizationally. Since $\delta$ satifies the strong triangle inequality, any indirect path through an intermediate permission $p_j$ can never exceed the largest direct gap. Hence the diameter truly reflects the worst-case separation in a single step, and by construction $\delta \in (0,1]$, so 
\[
0 < BlastRadius(P) \leq 1.
\]

A higher blast radius suggests that $P$'s access spans diverse or unrelated organizational zones, thus increasing the risk of data exfiltration or data forgery.

\medskip

\textbf{Theorem} \newline
Let $p_1, p_2, p_3$ be three permissions with associated scopes $s_1, s_2, s_3$ in a management group hierarchy. 
Define the blast radius distance by
\[
\delta(s_i, s_j) =
\begin{cases}
\displaystyle \frac{1}{2^{2d_{ij}+1}}, & \text{if } \mathrm{impact}(p_i, p_j) = 1, \\[1ex]
\displaystyle \frac{1}{2^{2d_{ij}}}, & \text{if } \mathrm{impact}(p_i, p_j) = 2,
\end{cases}
\]
where $d_{ij}$ is the depth of the lowest common ancestor of $s_i$ and $s_j$. Then $\delta$ satisfies the strong triangle inequality:
\[
\delta(s_1, s_3) \le \max\left\{ \delta(s_1, s_2),\, \delta(s_2, s_3) \right\}.
\]

\medskip

\textbf{Proof} \newline
\[
	D(s_i,s_j)\;=\;2^{-(2d_{ij}+1)} = (\frac{1}{2}) \cdot (\frac{1}{d_{ij}})^{2}
\]
is half the square of an ultrametric distance, hence is ultrametric (this is a well-known property of ultrametry):
\[
D(s_1,s_3)\;\le\;\max\{D(s_1,s_2),\,D(s_2,s_3)\}.
\]
By definition, $\delta$ can be written as:
\[
\delta(s_i,s_j)
=\kappa_{ij}\,D(s_i,s_j),
\quad
\kappa_{ij}\in\{1,2\}.
\]

\medskip
\noindent
Assume for contradiction that
\[
\delta(s_1,s_3)\;>\;\max\{\delta(s_1,s_2),\,\delta(s_2,s_3)\}.
\]
Write this as
\[
\kappa_{13}\,D(s_1,s_3)
\;>\;\max\{\kappa_{12}\,D(s_1,s_2),\,\kappa_{23}\,D(s_2,s_3)\}.
\]
Since each \(\kappa_{ij}\le2\), the right‐hand side is
\[
\max\{\kappa_{12}D(s_1,s_2),\kappa_{23}D(s_2,s_3)\}
\;\le\;
2 \,\max\{D(s_1,s_2),\,D(s_2,s_3)\}.
\]
Thus the assumption implies
\[
\kappa_{13}\,D(s_1,s_3)\;>\;2\,\max\{D(s_1,s_2),D(s_2,s_3)\}.
\]
We now split on two cases for \(\kappa_{13}\):

1. If \(\kappa_{13}=1\), then
   \[
   D(s_1,s_3) \;>\; 2\,\max\{D(s_1,s_2),D(s_2,s_3)\}
   \;\ge\;\max\{D(s_1,s_2),D(s_2,s_3)\},
   \]
   contradicting the ultrametric property of \(D\).

2. If \(\kappa_{13}=2\), then
   \[
   2\,D(s_1,s_3) \;>\; 2\,\max\{D(s_1,s_2),D(s_2,s_3)\}
   \;\Longrightarrow\;
   D(s_1,s_3) \;>\;\max\{D(s_1,s_2),D(s_2,s_3)\},
   \]
   again contradicting the ultrametric property of \(D\).

In both cases we reach a contradiction.  Therefore our assumption must be false, and
\[
\delta(s_1,s_3)\;\le\;\max\{\delta(s_1,s_2),\,\delta(s_2,s_3)\},
\]
showing that \(\delta\) is indeed an ultrametric.
\section{Conclusion}

This paper presents a dual-metric framework for scoring privilege strength and data risk in Microsoft Azure environments. 

We introduced the \emph{WAR distance}, superincreasing over Write, Action, and Read permissions, which enables a total ordering of principals by their control-plane privilege. 
The metric balances scope and operation type in a principled way, providing a foundation for de-escalation strategies, anomaly detection, and least-privilege governance.

\medskip

\noindent
Complementing this, we have introduced the \emph{Blast Radius}, an ultrametric-based metric for quantifying the maximum extent of Data Plane permissions in Azure cloud environments.
By emphasizing that both `Read' and `Write' dataActions are most critical, and by collapsing granularity below the subscription level, we capture the risk that a principal's permissions span disparate
organizational domains. This model complements control plane metrics such as WAR by providing a clear, mathematically robust measure of data-centric risks.

Together, these metrics form a coherent approach to reasoning about privilege in Azure—offering both mathematical rigor and practical applicability. This framework, implemented in the silhouette tool, enables structured comparisons, lifecycle monitoring, and risk quantification, they empower security teams to proactively reduce overprivilege, enforce least privilege, and build more resilient cloud infrastructures.

\medskip

\noindent

A note on the nature of scores: The WAR model produces a norm—induced by the WAR distance—between two principals, enabling ranking and comparative privilege analysis. In contrast, the data plane blast radius is computed over the permissions of a single principal. It is not a norm but a scalar measurement, derived from an ultrametric distance within that principal’s permission set. (And when this set is a singleton, the blast radius is not even derived from a distance). While the purpose of the blast radius is to compare and rank principals, the result is not a formal distance between them.

\subsection*{Future Work}
\begin{itemize}
	\item In the Control Plane, as of today Azure makes no clear distinction between pure configuration operations and IAM operations (role assignments). We plan to separate scoring IAM operations and configuration operations using a new, dedicated norm for the former 
and ignoring role management operations for the latter.
\item Replacing the max-based calculation of the WAR norm with a union-based calculation in the case of groups membership would ensure the decomposition of the norm into individual W, A and R coordinates remains possible.
\item The WAR norm of a single subscription-bound principal is the same as the WAR norm of a principal interacting with many subscriptions.  We plan to offer a better resolution for distinguishing both, since the latter is more akin to a MG-bound principal.
\end{itemize}

\newpage

\section{Appendix A: WAR Norm Examples}

\subsection{The WAR Norm Table}

\begin{figure}[htbp]
    \centering
    \includegraphics[width=0.45\textwidth]{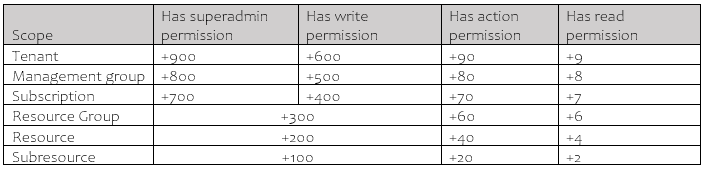}
    \caption{The WAR norm table}
    \label{fig:warnormtable}
\end{figure}

\subsection{Examples}
Here are a few examples of silhouette configurations based on the WAR norm table (by decreasing order of privileges):

\begin{enumerate}
	\item 999 corresponds to Tenant admin
	\item 888 corresponds to management group level superadmin
	\item 777 corresponds to subscription level superadmin
	\item 477 corresponds to subscription level for W, A and R
	\item 466 corresponds to subscription level for W, resource group level for A and R
	\item 377 corresponds to resource group level for W, subscription level for A and R
	\item 367 corresponds to resource group level for W and A, subscription level for R
	\item 346 corresponds to resource group level for W and R, resource level for A
	\item 026 corresponds to subresource level for A, resource group level for R, and no W action
	\item 000 corresponds to no control plane rights (except IAM roles management, as explained above)
\end{enumerate}

\newpage

\section{Appendix B: Blast Radius Examples}

\subsection{Single Permission At Management Group Level}

\noindent
\textbf{Scenario:} A principal is granted a single \texttt{Write} data plane permission scoped at the management group level. The management group resides at depth 3 in our hierarchical model (where \texttt{Tenant} is at depth 0, followed by \texttt{Management Groups}, and \texttt{Subscriptions} at the leaf level).

\medskip

\noindent
\textbf{Calculation:} With only one permission, there are no permission pairs to compare, we are in the singleton case.  The impact factor for a \texttt{Write} permission is 1. Since the permission resides within a single cluster at depth 3, the blast radius is:

\[
\text{BlastRadius}(P) = \frac{1}{2^{2 \cdot 3 + 1}} = 0.0078125
\]

\noindent
This value reflects localized motion, fully contained within the given management group hierarchy.

\subsection{Permissions Within The Same Management Group Branch}

\noindent
\textbf{Scenario:} A principal holds two data plane permissions:
\begin{itemize}
    \item A \texttt{Write} permission at a management group (depth 3), same as in the previous example.
    \item A \texttt{*} (wildcard) permission at a subscription under that same branch, located at depth 5.
\end{itemize}

\medskip

\noindent
\textbf{Calculation:} The LCA of both permissions is the original management group at depth 3. Although the wildcard permission implies a potentially higher impact (up to 2), the impact is computed \emph{at the shallowest depth}. Since only the \texttt{Write} permission is shallower than the wildcard, we retain the former hence the effective impact is 1.

\[
\text{BlastRadius}(P) = \frac{1}{2^{2 \cdot 3 + 1}} = 0.0078125
\]

\noindent
The blast radius remains unchanged compared to the single-permission case, as the additional permission does not widen the organizational scope beyond the original cluster.

\subsection{Permissions Spanning Different Management Group Branches}

\noindent
\textbf{Scenario:} The principal holds:
\begin{itemize}
    \item A \texttt{Write} permission in a management group at depth 3.
    \item A \texttt{Read} permission in a separate management group at depth 2
.
\end{itemize}

\noindent
The two permissions reside in distinct branches. Their LCA is a shared parent at depth 1.

\medskip

\noindent
\textbf{Calculation:} Both permissions have an impact factor of 1, and the LCA sits at depth 1:

\[
\text{BlastRadius}(P) = \frac{1}{2^{2 \cdot 1 + 1}} = 0.125
\]

\noindent
This significantly higher blast radius reflects the broader organizational spread of the principal's data access, signaling increased risk of data exfiltration or integrity compromise.

\subsection{Replacing Read Permission And Moving To The Tenant}
\textbf{Scenario:} A principal holds the usual `Write' permission at MG depth 3.  The principal has a `Read' permission under the Tenant (recall in our model, the Tenant sits at depth 0). 
The principal is then granted a second permission under Tenant scope, this time it is a `Write' permission.

\medskip

\noindent
\textbf{Calculation:} The principal has 3 permissions, but since the last two `Read' and `Write' are attached to the exact same scope, we collapse them into a `Write+Read' permission. Since it is at shallowest depth zero, and since it represents a high risk, its impact is 2. Finally, the LCA between MG and MG' is 0. So, 

\[
	\text{BlastRadius}(P)= \frac{2}{2^{2 \cdot 0 + 1}} = 1.0
\]

Interpretation: A blast radius of 1.0 is the maximum possible in our model. This extremely high score indicates that the principal's
permissions span the entire organizational boundary at the highest risk level for data exfiltration and forgery.

\end{document}